\documentclass[10pt]{elsarticle}

\RequirePackage{lineno}

\usepackage{graphicx}
\usepackage{subfiles}
\usepackage{color}
\usepackage{amsfonts}
\usepackage{amsmath}
\usepackage{amssymb}
\usepackage{wrapfig}
\usepackage{subfigure}
\usepackage[margin=0.75in,includefoot]{geometry}
\usepackage{url}
\usepackage[dvips]{epsfig}
\usepackage[footnotesize]{caption}  

\makeatletter
\def\ps@pprintTitle{%
 \let\@oddhead\@empty
 \let\@evenhead\@empty
 \def\@oddfoot{}%
 \let\@evenfoot\@oddfoot}
\makeatother

\begin{document}
\begin{frontmatter}
\title{PINGU Sensitivity to the Neutrino Mass Hierarchy\tnoteref{t1} }
\tnotetext[t1]{Submitted to the Snowmass 2013 Proceedings}
\author[Adelaide]{M.~G.~Aartsen} 
\author[MadisonPAC]{R.~Abbasi} 
\author[Gent]{Y.~Abdou} 
\author[Zeuthen]{M.~Ackermann} 
\author[Christchurch]{J.~Adams} 
\author[Geneva]{J.~A.~Aguilar} 
\author[MadisonPAC]{M.~Ahlers} 
\author[Berlin]{D.~Altmann} 
\author[MadisonPAC]{J.~Auffenberg} 
\author[Bartol]{X.~Bai\fnref{SouthDakota}} 
% ...
\address[Adelaide]{School of Chemistry \& Physics, University of Adelaide, Adelaide SA, 5005 Australia}
% ...
\address[Berlin]{Institut f\"ur Physik, Humboldt-Universit\"at zu Berlin, D-12489 Berlin, Germany}
% ...
\address[Christchurch]{Dept.~of Physics and Astronomy, University of Canterbury, Private Bag 4800, Christchurch, New Zealand}
% ...
\address[Geneva]{D\'epartement de physique nucl\'eaire et corpusculaire, Universit\'e de Gen\`eve, CH-1211 Gen\`eve, Switzerland}
\address[Gent]{Dept.~of Physics and Astronomy, University of Gent, B-9000 Gent, Belgium}
% ...
\address[MadisonPAC]{Dept.~of Physics and Wisconsin IceCube Particle Astrophysics Center, University of Wisconsin, Madison, WI 53706, USA}
% ...
\address[Bartol]{Bartol Research Institute and Department of Physics and Astronomy, University of Delaware, Newark, DE 19716, USA}
% ...
\address[Zeuthen]{DESY, D-15735 Zeuthen, Germany}
\cortext[cor1]{Corresponding authors}
\fntext[SouthDakota]{Physics Department, South Dakota School of Mines and Technology, Rapid City, SD 57701, USA}
\fntext[LosAlamos]{Los Alamos National Laboratory, Los Alamos, NM 87545, USA}
\fntext[Bari]{also Sezione INFN, Dipartimento di Fisica, I-70126, Bari, Italy}
\fntext[Goddard]{NASA Goddard Space Flight Center, Greenbelt, MD 20771, USA}

\begin{abstract}
\noindent
The neutrino mass hierarchy is one of the few remaining unknown
parameters in the neutrino sector and hence a primary focus of the
experimental community.  The Precision IceCube Next Generation Upgrade (PINGU)
experiment, to be co-located with the IceCube DeepCore detector in the
deep Antarctic glacier, is being designed to provide a first
definitive measurement of the mass hierarchy.  We have conducted
feasibility studies for the detector design that demonstrate a
statistically-limited sensitivity to the hierarchy of 2.1$\sigma$ to
3.4$\sigma$ per year is possible, depending on the detector geometry (20 to 40 strings) and analysis efficiencies. First
studies of the effects of systematic and theoretical uncertainties show
limited impact on the overall sensitivity to the hierarchy.  Assuming
deployment of the first array elements in the 2016/17 austral summer
season a 3$\sigma$ measurement of the hierarchy is anticipated with
PINGU in 2020.
\end{abstract}
\end{frontmatter}

\noindent
The mixing angles and mass-squared differences that describe
oscillations in the neutrino sector have been measured with high
precision through the efforts of a variety of experiments worldwide~\cite{PDG_review}. We plan to submit a proposal to build the Precision IceCube Next
Generation Upgrade (PINGU) experiment to measure one of the few
remaining unknowns in the neutrino sector, the neutrino mass hierarchy
(NMH). PINGU will leverage the demonstrated ability of IceCube's
DeepCore in-fill array to measure atmospheric neutrino oscillation
parameters~\cite{IceCubeosc} via deployment of a further increase in the photon detector density
in the DeepCore region. This will enable us to isolate and reconstruct
a high-statistics sample of 5--20 GeV atmospheric muon neutrinos that
undergo matter effects over a wide range of baselines, providing
sensitivity to the NMH~\cite{Mena,ARS,Winter}. Based on our preliminary estimates, we
expect that we could determine the hierarchy with a statistical
significance of 3$\sigma$ with 1--2 years of data, and a significance
of 5$\sigma$ using 2--4 additional years of data.

The PINGU design and construction follows closely that of IceCube,
with similar hot-water drilling techniques, down-hole cables,
deployment strategy, Digital Optical Module (DOM) hardware, and online
and offline software. We are evaluating the NMH sensitivity of three
distinct detector geometries with either 20 or 40 strings of DOMs, and
60 to 100 DOMs per string, chosen to bracket the range of geometries
deployable in either a two- or three-year period and consistent with
drilling and deployment constraints. Using well-established metrics
from IceCube experience, the cost to construct and deploy PINGU is
estimated to be between \$8M--12M in startup costs associated with
setting up DOM assembly and refurbishing the hot water drill used for
IceCube construction, presuming the existing drill equipment remains
available for use by PINGU, plus roughly \$1.25M per string related to
the detector hardware and deployment. We anticipate that some of this
funding, as well as logistical support, would come from US funding
sources, but that a considerable portion would be provided by
international partners. First deployments could start as early as the
2016/17 South Pole summer season, and be completed in 2--3 seasons
depending on geometry.  This would permit a determination of the
hierarchy with $3\sigma$ significance by 2020.

Detailed studies of the performance of PINGU and the significance with
which it would determine the NMH are ongoing. These studies address
detector energy and angular resolution, background rejection,
systematic uncertainties, and the impact of degeneracies with physics
parameters other than the NMH.  At present, three independent
estimates of the PINGU sensitivity to the NMH have been developed
using different statistical techniques and assumptions regarding
detector performance and including different combinations of physics
degeneracies and detector systematics.  Each study was designed to
evaluate the impact of a particular factor or group of factors which
may impact PINGU's sensitivity, as discussed below. While we continue
to work to include the full details in a single complete study, these
targeted investigations give us confidence that there are no
fundamental problems that could prevent a measurement of the NMH with
PINGU within a few years.  Since this work is still in progress, we
present a range of estimated sensitivity (see figure), presenting both
the different geometries under study as well as a range of predicted
performance of background rejection and flavor identification
algorithms. 
\begin{wrapfigure}{r}{12cm}
 \includegraphics[width=12cm, angle=0]{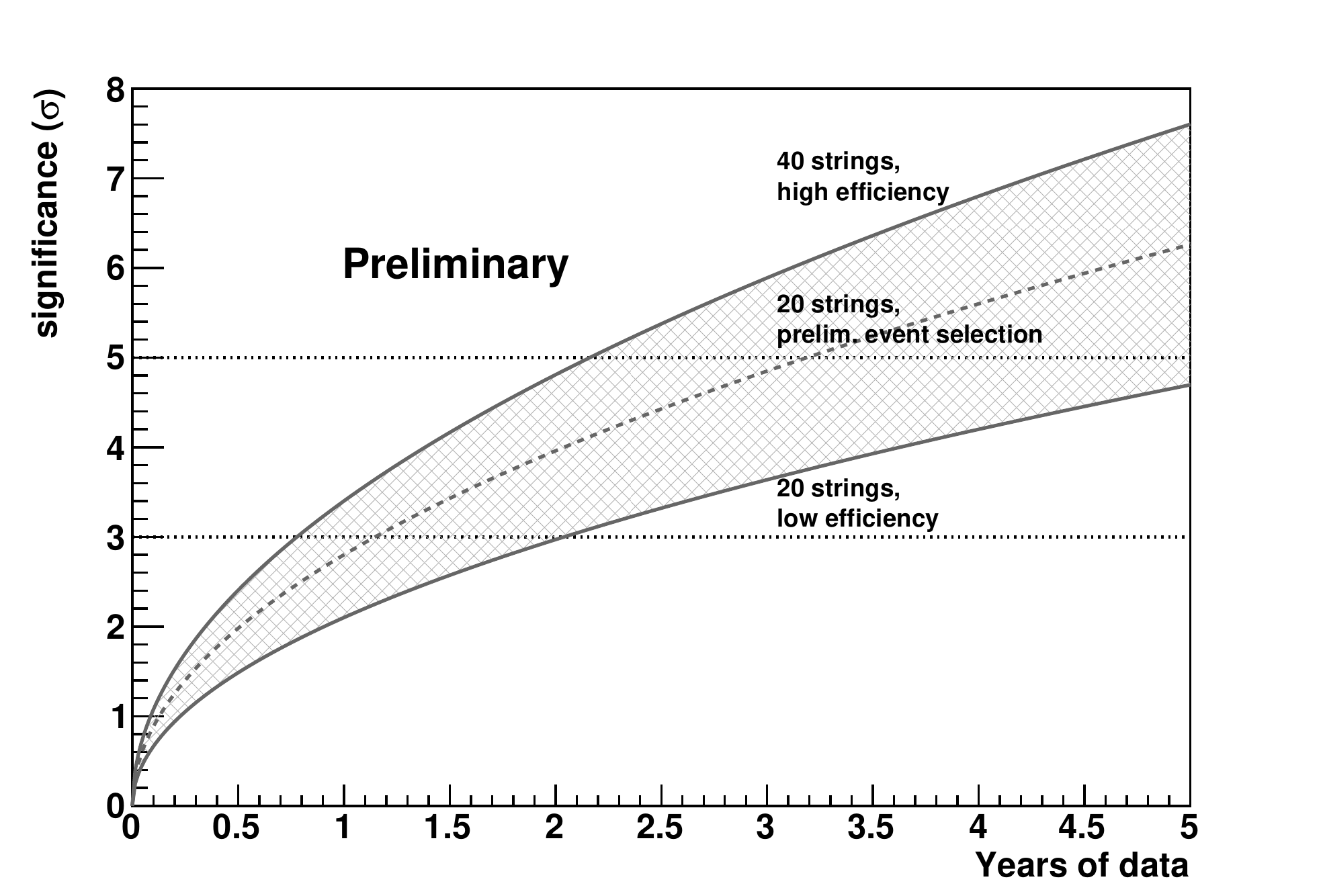}
 {Estimated significance for determining the neutrino mass hierarchy
   with PINGU. The top of the range is based on a 40 string detector
   with a high assumed signal efficiency in the final analysis; the
   bottom uses a 20 string detector and assumed a lower signal
   efficiency.}
     \label{fig:nmhsensitivity}
\end{wrapfigure}
From the preliminary studies we anticipate the sensitivity
will be statistically limited for the first few years, and the
projected time dependence is based on this assumption. All estimates
are based on a full three-flavor treatment of neutrino oscillations
including matter effects, and although to date we have not fully
examined the influence of non-zero $\delta_\mathrm{CP}$
or uncertainties in the true values of $\theta_{12}$ and
$\Delta\mathrm{m}^2_{12}$, their impact is not expected to be large
[4]. We have found that the ability to measure neutrino oscillations
across a wide range of energies and baselines allows us to constrain
our systematic uncertainties and resolve degeneracies between physics
parameters quite accurately from the data itself. 
The matter effects
we would exploit to measure the NMH produce distinctive signatures in
the energy-angle space, and systematic uncertainties related to the
detector do not reproduce these complicated patterns. We note that an
additional study using a simplified detector model was conducted to
assess the relative importance of physics degeneracies, reaching
conclusions similar to~\cite{ARS,Winter}, and suggesting that a sensitivity of
5$\sigma$ could be reached before systematic effects limit further
increases in sensitivity.  

Event quality and selection are key elements in these studies.  The
simulations of the angular and energy resolution of the three detector
geometries have been conducted using established, computationally fast,
DeepCore algorithms optimized for the PINGU geometry. These algorithms
yield a median neutrino energy resolution of about (0.7~GeV +
0.2$\times$E$_{\nu}$), and a median neutrino angular resolution
improving from 15$^{\circ}$ to 8$^\circ$ as E$_{\nu}$ increases from 5
to 20~GeV. More computationally intensive algorithms yield better
resolutions at higher efficiency, but we use the fast algorithms in
the studies presented here of PINGU's NMH sensitivity, partly to be
conservative and partly to reduce turnaround time while studying the
systematic uncertainties. We assume that we will be able to reduce the
atmospheric muon background rate to a low level without substantial
loss of signal efficiency based on our experience with DeepCore~\cite{IceCubeosc,IceCubecasc}
and on the knowledge that PINGU will benefit from the enhanced active
vetoing provided by the outermost DeepCore strings. Studies of
atmospheric muon rejection are underway to confirm this
assumption. After reducing the atmospheric muon background, we expect
that neutrino events other than $\nu_{\mu}$ CC will dominate as the
remaining background.  The three estimates discussed below use
different methods to estimate the effect of this background.

The first analysis models a 40 string detector, and makes aggressive
assumptions regarding signal efficiency.  Our more computationally
expensive event reconstructions indicate that approximately 85\% of
neutrino-induced shower events reconstruct with a track length L$_\mu
<15$~m; rejecting events reconstructing with shorter L$_\mu$
corresponds to imposing a muon energy threshold of $\sim3$~GeV and a
neutrino energy threshold of $\mathrm{E}_{\nu}\sim6$~GeV, which is
comparable to the threshold for successful event reconstruction with
the faster algorithms used in this study.  In this first analysis, we
therefore assume that this efficiency of successful event
reconstruction will be comparable to the final efficiency of the
eventual PINGU analysis.

The analysis uses a binned $\chi^{2}$ approach using ``pull factors'' 
to account for experimental and theoretical systematic
uncertainties~\cite{Fogli}. The simulated data sets used for the analysis are
reconstructed using the fast algorithms discussed above, binned in the
reconstructed energy and angle, with bin widths commensurate with the
expected resolutions.  Systematic uncertainties are incorporated into
the analysis as nuisance parameters in the $\chi^{2}$ sum and are
simultaneously fit to the data with penalty terms according to the
current estimated uncertainties.  An approximation for the median
sensitivity of the detector is provided by the analysis of a
representative dataset, also known as the ``Asimov'' data set~\cite{Cowan},
under the assumption that the test statistic will be approximately
$\chi^{2}$ distributed despite the discrete nature of the
measurement. We verified the accuracy of this procedure using
ensembles of pseudo-experiments in which both the mass splitting and
the values of the nuisance parameters responsible for the primary
systematic uncertainties were fit, and found that the approximation
was generally conservative. We have evaluated the impact of systematic
uncertainties in our effective volume for neutrino events (30\%) and
in the spectral index of the atmospheric muon neutrino flux ($\pm
0.05$), as well as potential biases in our absolute energy scale
calibration (10\%) and directional reconstruction (10\%), and errors
in our estimated energy and angular resolutions (10\%). Theoretical
uncertainties in the values of $\theta_{23}$ and
$\Delta\mathrm{m}^{2}_\mathrm{atm}$ are treated in the same
manner. With these systematics included, we obtain a median expected
significance of 3.4$\sigma$ per year for the current world average
values of the oscillation parameters (top curve in figure).  Considering the full range of atmospheric
mass splitting and mixing angle preferred by current world measurements, a significance of 2.8$\sigma$ or
better is expected from one year of data.

A second analysis models a less densely instrumented 20 string
detector and uses a similar statistical analysis, although a
likelihood ratio is calculated rather than a $\chi^{2}$.  This
analysis includes full event reconstruction and background rejection
based on DeepCore analysis methods, modified for the denser PINGU
geometry and using the outermost DeepCore strings to veto atmospheric
muons in view of the smaller PINGU fiducial volume.  Signal
efficiencies of around 9\% are obtained in the range from 5--15~GeV.
While this event selection provides some rejection of NC and $\nu_{e}$
backgrounds, the residual backgrounds are accommodated in the
statistical analysis.  The $\nu_{e}$ flux normalization is treated as
an additional nuisance parameter to accommodate uncertainties in the
production rate in air showers.  This analysis included the
theoretical uncertainty on $\theta_{13}$ as well as the atmospheric
mass splitting and mixing angle, but did not examine the impact of
potential reconstruction biases or errors in detector resolution.
This analysis found an expected significance of 2.8$\sigma$ from one
year of PINGU data (middle curve in figure) assuming the world average oscillation parameters
(including non-zero $\delta_\mathrm{CP}$) in the inverted hierarchy
scenario, and slightly higher for the normal hierarchy.

The third analysis used the more CPU-intensive approach of generating
and fitting ensembles of pseudo-experiments using a likelihood
ratio-based analysis.  This analysis also considered the 20-string
detector but assumed that a much lower signal efficiency, ranging from
about 1\% at 5~GeV to 30\% at 15~GeV, would be achieved after
rejection of backgrounds due to atmospheric muon and neutrino-induced
cascades. To reduce processing time, detector resolution effects were
modeled with a Gaussian smearing based on the resolutions given above
instead of explicit reconstruction of each event.  Even with the lower
assumed efficiencies used in this analysis, we estimate that a
3$\sigma$ measurement of the hierarchy would be possible with 2 years
of data (bottom curve in figure). Given the shorter deployment schedule of the smaller
detector, this measurement could also be completed by 2020.  Due to
the computational intensity of this approach only the primary physics
systematic, which was found to be the value of
$\Delta\mathrm{m}^2_{31}$, consistent with~\cite{ARS,Winter}, was explicitly
included in this study.  However, just as in the first analysis
described above, we find that the expected significance of the
measurement is essentially independent of the true value of
$\Delta\mathrm{m}^2_{31}$ when the full energy-angle range of the data
is used to determine the NMH. Further work is underway to confirm that
the other systematics (including the mixing angles and
$\delta_{\mathrm{CP}}$) can be similarly handled without reducing the
expected significance.

While we continue to work to optimize the PINGU geometry, to model
more accurately the impact of background rejection and physics
degeneracies on the final analysis, and to investigate further the
impact of possible detector-related systematics, we believe that these
studies demonstrate that all of the major potential issues are
manageable and that a measurement of the NMH would be possible with
PINGU by 2020. We are currently preparing a detailed Letter of Intent
discussing these studies and secondary PINGU physics topics, to be
followed by a full proposal.

\end{document}